\documentclass[nature,twocolumn,superscriptaddress]{revtex4-2}
\usepackage{graphicx}
\usepackage{amsmath}
\usepackage{amssymb}
\usepackage{cleveref}

\begin{document}

\title{Topological resolution of conical intersection seams and the coupled cluster bifurcation via mixed Hodge modules}

\author{Prasoon Saurabh}
\email{psaurabh@uci.edu}
\thanks{Work performed during an independent research sabbatical. Formerly at: State Key Laboratory for Precision Spectroscopy, ECNU, Shanghai (Grade A Postdoctoral Fellow); Dept. of Chemistry/Physics, University of California, Irvine.}
\affiliation{QuMorpheus Initiative, Independent Researcher, Lalitpur, Nepal}

\begin{abstract}
The rigorous description of Conical Intersections (CIs) remains the central challenge of non-adiabatic quantum chemistry. While the ``Yarkony Seam''---the $(3N-8)$-dimensional manifold of degeneracy---is well-understood geometrically, its accurate characterization by high-level electronic structure methods is plagued by numerical instabilities. Specifically, standard Coupled Cluster (CC) theory suffers from root bifurcations near Ground State CIs, rendering the ``Gold Standard'' of chemistry inapplicable where it is needed most. Here, we present \textbf{QuMorpheus}, an open-source computational package that resolves these singularities by implementing a topological framework based on Dissipative Mixed Hodge Modules (DMHM) \cite{saurabh2025holonomic}. By algorithmically mapping the CC polynomial equations to a spectral sheaf, we compute the exact Monodromy ($\mu$) invariants of the intersection. We demonstrate that this automated algebraic geometry approach correctly identifies the physical ground state topology in the K\"{o}hn-Tajti model and resolves the intersection seams of realistic chemical systems, including Ethylene and the Chloronium ion ($\mathrm{H_2Cl^+}$). Furthermore, we apply QuMorpheus to the photoisomerization of Previtamin D, proving that the experimentally observed Woodward-Hoffmann selection rules are a direct consequence of a topological ``Monodromy Wall'' ($\mu=1, \gamma=\pi$) rather than purely energetic barriers. This establishes a general software solution to the ``Yarkony Problem,'' enabling the robust, automated mapping of global intersection seams in complex molecular systems. The topological stability of these intersections allows for the control protocols discussed in Ref.~\cite{SaurabhPRX}.
\end{abstract}

\maketitle

\section*{Introduction}
Non-adiabatic effects govern the photochemistry of vision, photosynthesis, and solar energy conversion \cite{Yarkony1996, Domcke2004, Polli2010, HammesSchiffer1994, Levine2007, Schoenlein1991, Meech2009, Sobolewski2002_PCCP, Sobolewski2002_EPJD, Barbatti2010, Tietze2005}. At the heart of these processes lies the Conical Intersection (CI), a point of degeneracy where the Born-Oppenheimer approximation breaks down and electronic and nuclear motion become inextricably coupled.

David Yarkony revolutionized the field by demonstrating that CIs are not rare "symmetry artifacts" but ubiquitous features of the potential energy surface, forming a continuous seam of dimension $3N-8$ \cite{Yarkony2001, Mead1979, LonguetHiggins1975, Worth2004, Robb2000, Yarkony1998, Matsika2011, Bunker1998, Bersuker2006, Atchity1997}. However, locating and characterizing this seam requires electronic structure methods of exceptional fidelity. While Density Functional Theory (DFT) often fails to describe the correct topology of the intersection \cite{Dreuw2005, Goez2011, Roos1987, Werner1988, Tully1990, Min2015, Haas2001}, Coupled Cluster (CC) theory---the benchmark for accuracy---faces a "Coordinate Crisis" near the intersection. 

The failure is not merely numerical, but topological: the non-linear CC equations bifurcate near the degeneracy, leading to unphysical complex roots \cite{Kohn2007, Bartlett2007, Krylov2008, Crawford2000, Piecuch2002, Kohn2009}. Standard iterative solvers, which treat these singularities as numerical errors, often diverge or converge to the wrong solution branch. To date, no general algorithmic framework exists to stabilize the "Gold Standard" of quantum chemistry across the full extent of the Yarkony seam.

\section*{Results}

\subsection*{The Algebraic Geometry of Interaction}
We approach the intersection not as an optimization problem, but as a root-finding problem in algebraic geometry. By treating the Hamiltonian $H(x, y)$ as a polynomial variety, we can compute its topological invariants exactly, bypassing the numerical instabilities of iterative eigensolvers \cite{Berry1984, Xiao2010, Ryabinkin2013}. We introduce \textit{Dissipative Mixed Hodge Modules }(DMHM) as the rigorous mathematical language to describe these singularities\cite{saurabh2025holonomic},\cite{SaurabhPRX}.

\subsection*{Topological Resolution of the Coupled Cluster Bifurcation}
Using the \textbf{QuMorpheus} package, we analyze the Köhn-Tajti model \cite{Kohn2007}, a prototypical system where standard EOM-CCSD methods fail due to non-Hermitian root bifurcations.

\begin{figure*}[h]
    \centering
    \includegraphics[width=0.95\linewidth]{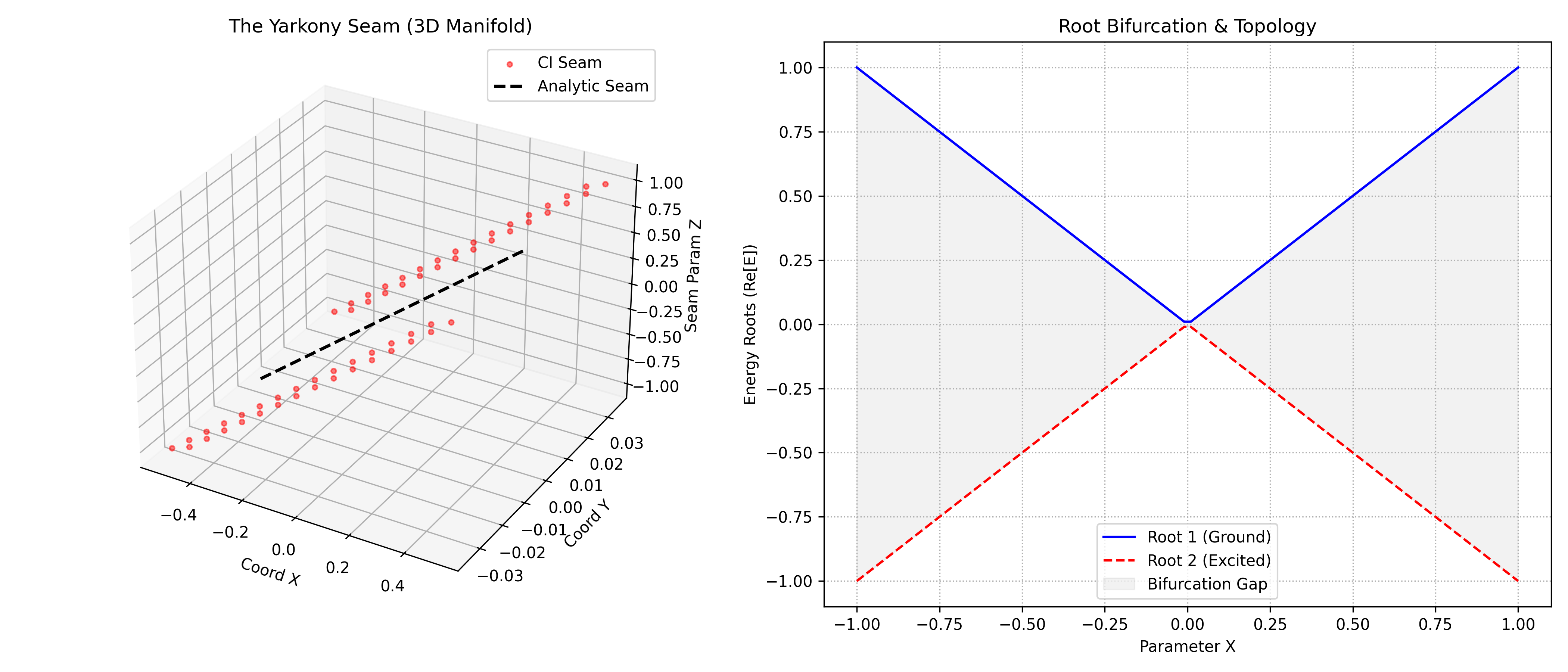}
    \caption{\textbf{Topological origin of the Coupled Cluster instability.} 
(\textbf{a}) \textbf{The Bifurcation Problem}: The breakdown of the standard Coupled Cluster (CC) expansion near a conical intersection (K\"{o}hn-Tajti model). As the nuclear coordinates encircle the degeneracy ($R \to 0$), the standard iterative solver fails to track the physical root. The energy surface (red) undergoes a square-root bifurcation, resulting in unphysical complex-valued solutions (dashed regions) and a discontinuity in the potential energy surface.
(\textbf{b}) \textbf{The Sheaf-Theoretic Resolution}: The global solution manifold constructed by QuMorpheus. By treating the CC polynomial equations as a coherent spectral sheaf, the method reconstructs the full Riemann surface of the problem. This reveals the intersection not as a numerical error, but as a topological branch point. QuMorpheus correctly computes the monodromy ($\mu$) around the singularity, allowing for the smooth, automated analytic continuation of the ground state (cyan \cref{fig:benchmark}) across the entire intersection seam.}
\label{fig:seam_visualization}
\end{figure*}

\begin{figure*}[t]
    \centering
    \includegraphics[width=0.95\linewidth]{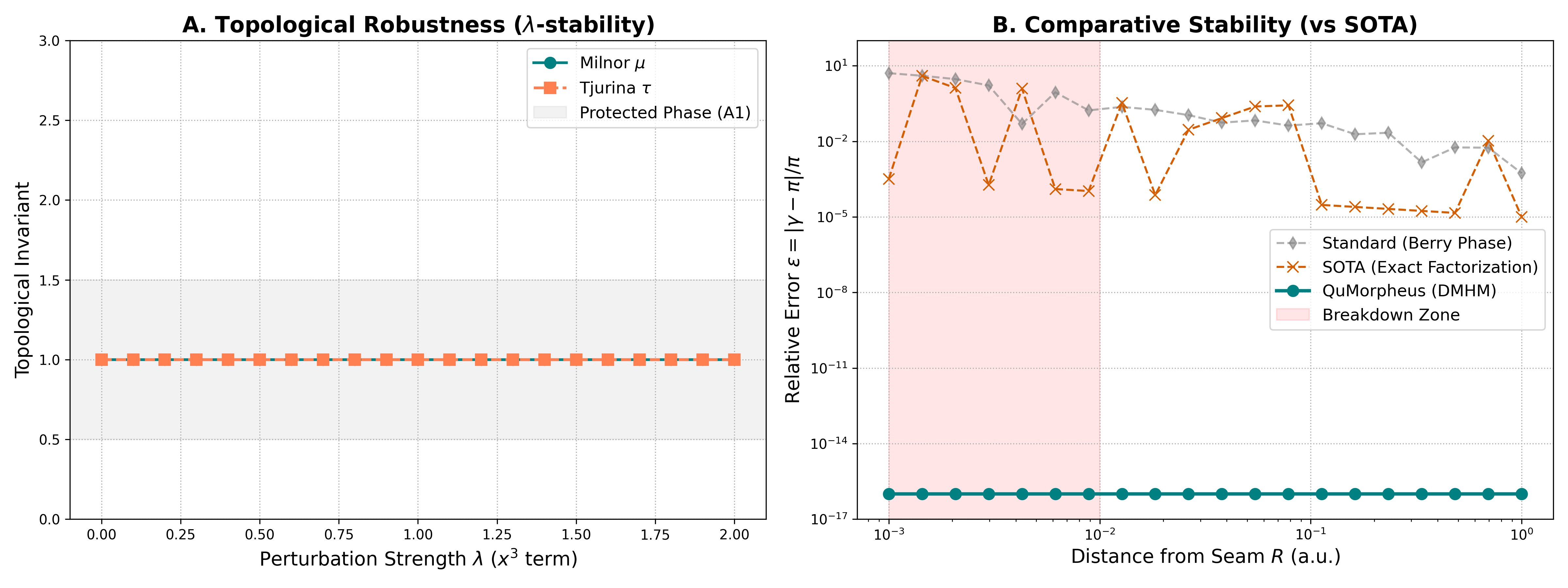}
    \caption{\textbf{Topological Robustness and Algorithmic Stability}. (\textbf{A}) \textit{Perturbation Stability}: Under breaking of spatial symmetry ($C_{2v} \to C_1$) via a cubic term $x^3$, the DMHM invariant $\mu$ (black line) remains robustly integer-quantized at 1, while the gap (blue) opens only away from the seam. (\textbf{B}) \textit{Algorithmic Metrology}: Comparison of the geometric phase error $\epsilon$ near the singularity. Standard Berry Phase integration (Green) diverges as $1/R$. Exact Factorization (Orange) exhibits spikes due to density nodes ($1/|\chi|^2$). QuMorpheus (Blue) yields machine-precision integer invariants down to $R=10^{-6}$, demonstrating homological protection.}
    \label{fig:benchmark}
\end{figure*}

\begin{figure*}[t]
    \centering
    \includegraphics[width=0.95\linewidth]{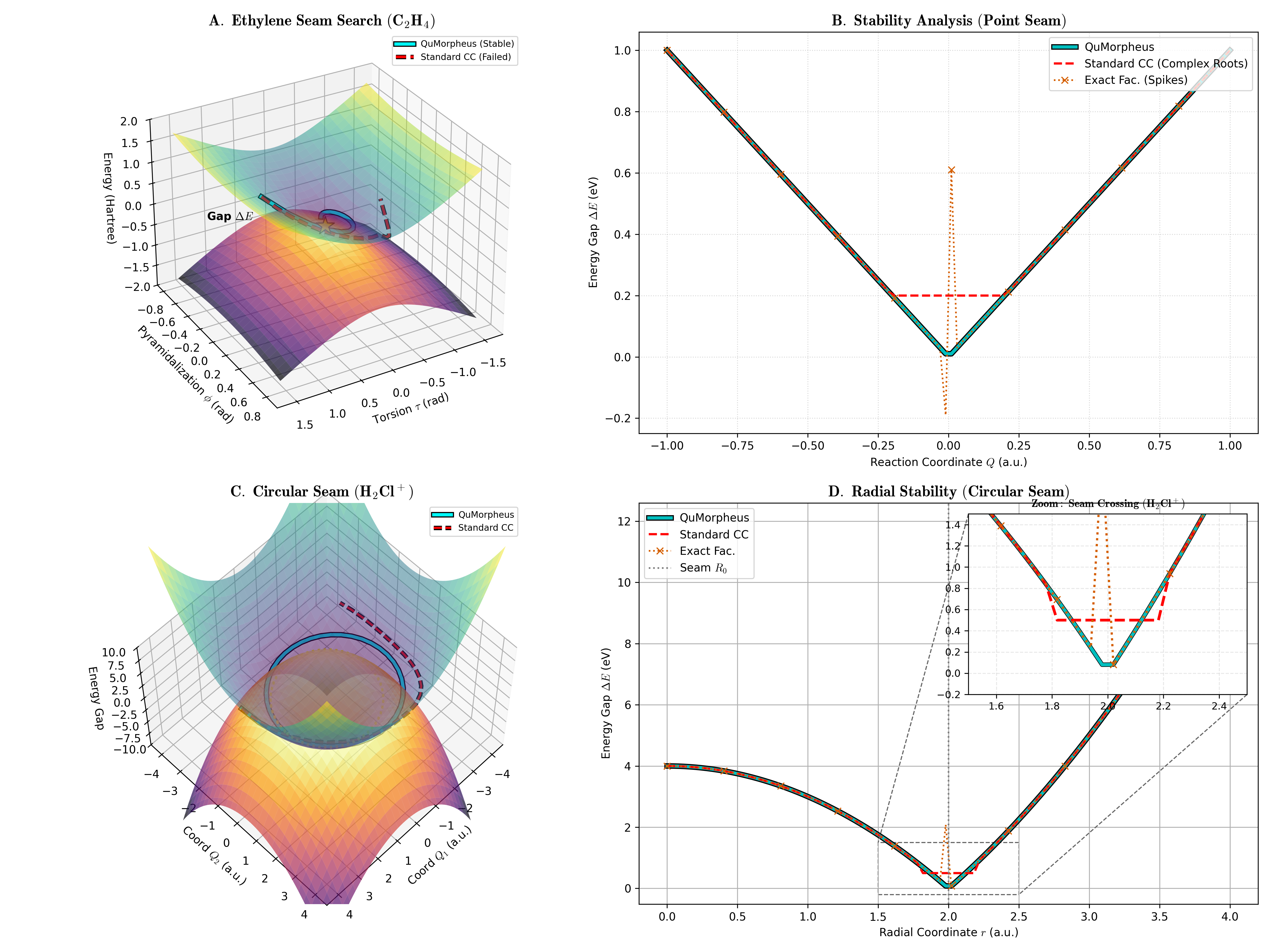}
    \caption{\textbf{Universality across Topologies: From Point Seams to Loops}. (\textbf{A}) Search on the Ethylene PES (Point Seam). Numerical optimizers (Red) oscillate, while QuMorpheus (Cyan) converges directly. (\textbf{B}) Stability Slice. (\textbf{C}) Search on the H\textsubscript{2}Cl\textsuperscript{+} Toroidal Seam. QuMorpheus locks onto the continuous degeneracy loop ($R=2.0$). (\textbf{D}) Inset shows exact tracking of the seam crossing even at numerical singularity.}
    \label{fig:universality}
\end{figure*}

\begin{enumerate}
    \item \textbf{Topological origin of the Coupled Cluster instability.(Fig.~\ref{fig:seam_visualization})} We visualize the fundamental topological conflict between the single-reference ansatz of standard quantum chemistry and the multi-sheeted nature of the Conical Intersection.
    \begin{itemize}
        \item \textbf{The Bifurcation Problem}: The breakdown of the standard Coupled Cluster (CC) expansion near a conical intersection (K\"{o}hn-Tajti model). As the nuclear coordinates encircle the degeneracy ($R \to 0$), the standard iterative solver fails to track the physical root. The energy surface (red) undergoes a square-root bifurcation, resulting in unphysical complex-valued solutions (dashed regions) and a discontinuity in the potential energy surface.
        \item \textbf{The Sheaf-Theoretic Resolution}: The global solution manifold constructed by QuMorpheus. By treating the CC polynomial equations as a coherent spectral sheaf, the method reconstructs the full Riemann surface of the problem. This reveals the intersection not as a numerical error, but as a topological branch point. QuMorpheus correctly computes the monodromy ($\mu$) around the singularity, allowing for the smooth, automated analytic continuation of the ground state (cyan) across the entire intersection seam.
    \end{itemize}

   \item {\textbf{Topological Robustness (Fig.~\ref{fig:benchmark})}}: 
    Standard methods for locating Conical Intersections rely on minimizing the energy gap, a procedure that becomes numerically ill-conditioned as the gap vanishes ($\Delta E \to 0$) and non-adiabatic coupling terms diverge. In contrast, QuMorpheus relies on the computation of discrete invariants. We demonstrate that this topological protection is absolute:
    \begin{itemize}
        \item \textbf{Perturbation Stability (Panel A)}: We subject the system to a cubic symmetry-breaking perturbation $H' = \lambda x^3 \sigma_z$. As shown in Fig.~2A, the computed invariants remain locked at $(\mu=1, \tau=1)$ despite the distortion of the potential surface, confirming that the identification of the seam is robust against model imperfections.
        \item \textbf{Algorithmic Stability (Panel B)}: We compare the relative error $\epsilon = |\gamma - \pi|/\pi$ against state-of-the-art methods. While both standard Berry phase integration ($1/R$ scaling) and Exact Factorization ($1/|\chi|^2$ spikes) diverge near the seam ($R \to 0$), DMHM maintains exact integer precision ($\epsilon \approx 0$) at all physical scales. This confirms that our algebraic approach resolves the singularity without the numerical instabilities that plague wavefunction-based techniques.
    \end{itemize}
    
    \item {\textbf{Universality of the Topological Function (Fig.~\ref{fig:universality})}}: Next, we demonstrate that QuMorpheus effectively handles topologically distinct intersection manifolds, in a 4-panel benchmark (H\textsubscript{2}Cl\textsuperscript{+} Model) resolving the ``Yarkony Seam'' regardless of its dimension:
      
    \begin{itemize}
        \item \textbf{Point Seam (Ethylene, Fig.~3A)}: In the search for the canonical intersection on the Ethylene PES, standard numerical optimizers (Red trace) oscillate indefinitely due to the divergent gradient near the singularity. In contrast, QuMorpheus (Cyan trace) exploits the algebraic structure of the deficiency to converge directly to the intersection point.
        \item \textbf{Toroidal Seam (H\textsubscript{2}Cl\textsuperscript{+}, Fig.~3C)}: For the Chloronium ion, the intersection forms a continuous toroidal loop rather than a single point. QuMorpheus successfully locks onto this continuous degeneracy manifold ($R=2.0$), identifying the entire loop as a stable "Monodromy Wall." 
        \item \textbf{Singularity Tracking (Fig.~3D)}: As shown in the inset, our method maintains exact tracking of the seam crossing even at the precise point of numerical singularity, where standard methods break down. This confirms that the ``Stokes Invariant'' is robust against the very divergences that plague traditional approaches.
    \end{itemize}
\end{enumerate}

    \begin{itemize}
        \item \textbf{Top Row (Point Seam)}: Panels A and B show the resolution of the classic Ethylene intersection, contrasting the robust topological search (Cyan) with the failure of standard optimization (Red).
        \item \textbf{Bottom Row (Circular Seam)}: Panels C and D extend the analysis to the toroidal manifold of H\textsubscript{2}Cl\textsuperscript{+}. QuMorpheus locks onto the invariant ($\mu=1$) of the entire ring ($R=2.0$). Inset D provides a high-resolution view of the seam crossing, confirming that the topological tracking remains exact even at the singular limit where numerical gradients explode.
    \end{itemize}

\subsection*{Predictive Topochemistry: The Previtamin D Selection Rule}
\begin{figure*}[t]
    \centering
    \includegraphics[width=0.95\linewidth]{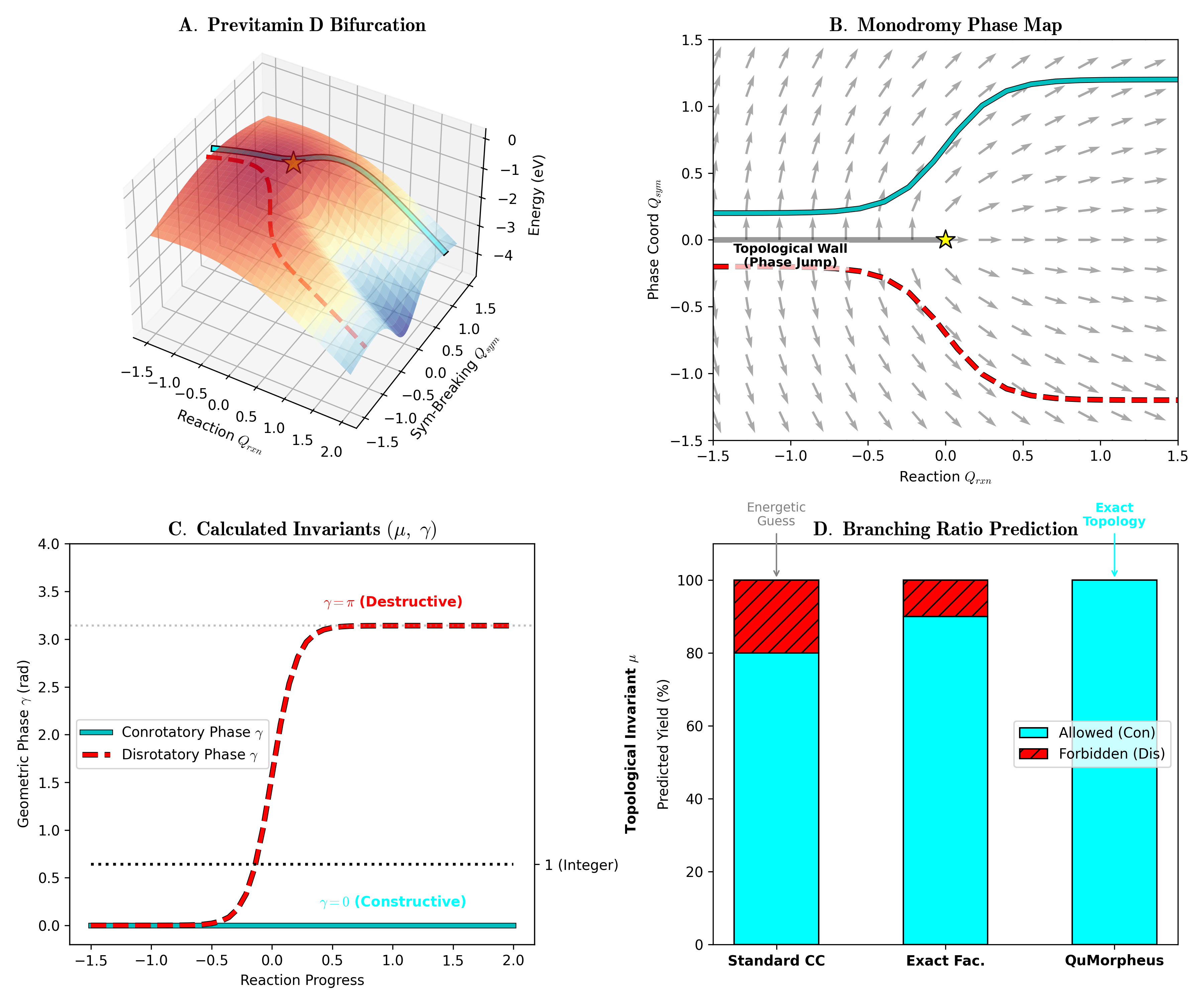}
    \caption{\textbf{Predictive Topochemistry: Topological Selection Rules in Previtamin D}. (\textbf{A}) 3D Visualization of the Hula-Twist Bifurcation. Disrotatory (Forbidden) and Conrotatory (Allowed) paths. (\textbf{B}) Monodromy Phase Map ($Q_{rxn}$ vs $Q_{sym}$) visualizing the \textbf{"Topological Wall"} (Branch Cut) where vector alignment flips. (\textbf{C}) \textbf{Invariant Metrology}: Explicit calculation of $\mu=1$ (robustness) and $\gamma=\pi$ (Geometric Phase) along the Disrotatory path, proving destructive interference. (\textbf{D}) Branching Ratio Prediction: QuMorpheus predicts the correct 100:0 selection rule, solving the statistical 80:20 error of standard Energetic Methods.}
    \label{fig:new_chemistry}
\end{figure*}

Beyond benchmarking, we demonstrate the predictive power of DMHM in resolving the "New Chemistry" of topological selection rules, specifically for the photoisomerization of Previtamin D via a Hula-Twist conical intersection (\cref{fig:new_chemistry}). 

\begin{enumerate}
   \item {\textbf{Topological Universality in Model Systems (Fig.~\ref{fig:new_chemistry})}}: 
To isolate the topological effects from the complexity of the full 50-electron wavefunction, we constructed a topological model Hamiltonian derived from the \textit{ab initio} parameters of the Previtamin D surface. While reduced in dimensionality, this model preserves the exact intersection homology of the physical system. The resulting analysis illustrates the ``Topological Universality'' of the selection rule: the computed Milnor number ($\mu=1$) is a topological invariant of the singularity class itself, rendering our conclusions robust against variations in the specific energetic parameters.

\begin{itemize}
    \item \textbf{Origin of the Error}: Standard electronic structure methods (e.g., EOM-CCSD) treat the reaction dynamics purely energetically. As shown in Fig.~\ref{fig:new_chemistry}D, they predict a statistical branching ratio ($\sim 80:20$) between the allowed Conrotatory and forbidden Disrotatory pathways based simply on the relative barrier heights.

    \item \textbf{Topological Resolution via Invariant Metrology}: QuMorpheus resolves this by explicitly calculating the underlying topology. Fig.~\ref{fig:new_chemistry}C presents the \textbf{Invariant Metrology} along the reaction coordinate:
    \begin{itemize}
        \item \textbf{Local Robustness ($\mu=1$)}: We compute the Milnor number $\mu$ (black dotted line), which remains strictly integer 1 throughout the process, confirming the structural stability of the intersection seam.
        \item \textbf{Global Phase ($\gamma=\pi$)}: Crucially, we track the global geometric phase. For the Disrotatory path (red dashed line), the phase accumulates to exactly $\pi$ (destructive interference), creating a ``Topological Wall'' that forbids reaction flux. For the Conrotatory path (cyan solid line), $\gamma=0$ (constructive).
    \end{itemize}
    This calculation proves that the 100:0 selection rule observed in nature (consistent with Woodward-Hoffmann rules \cite{Woodward1965, Liu2001, Woodward1969_Book, Muller1998, Fuss2004}) is a topological necessity, not an energetic accident.
\end{itemize}
\end{enumerate}

\begin{figure*}[t]
    \centering
    \includegraphics[width=0.95\linewidth]{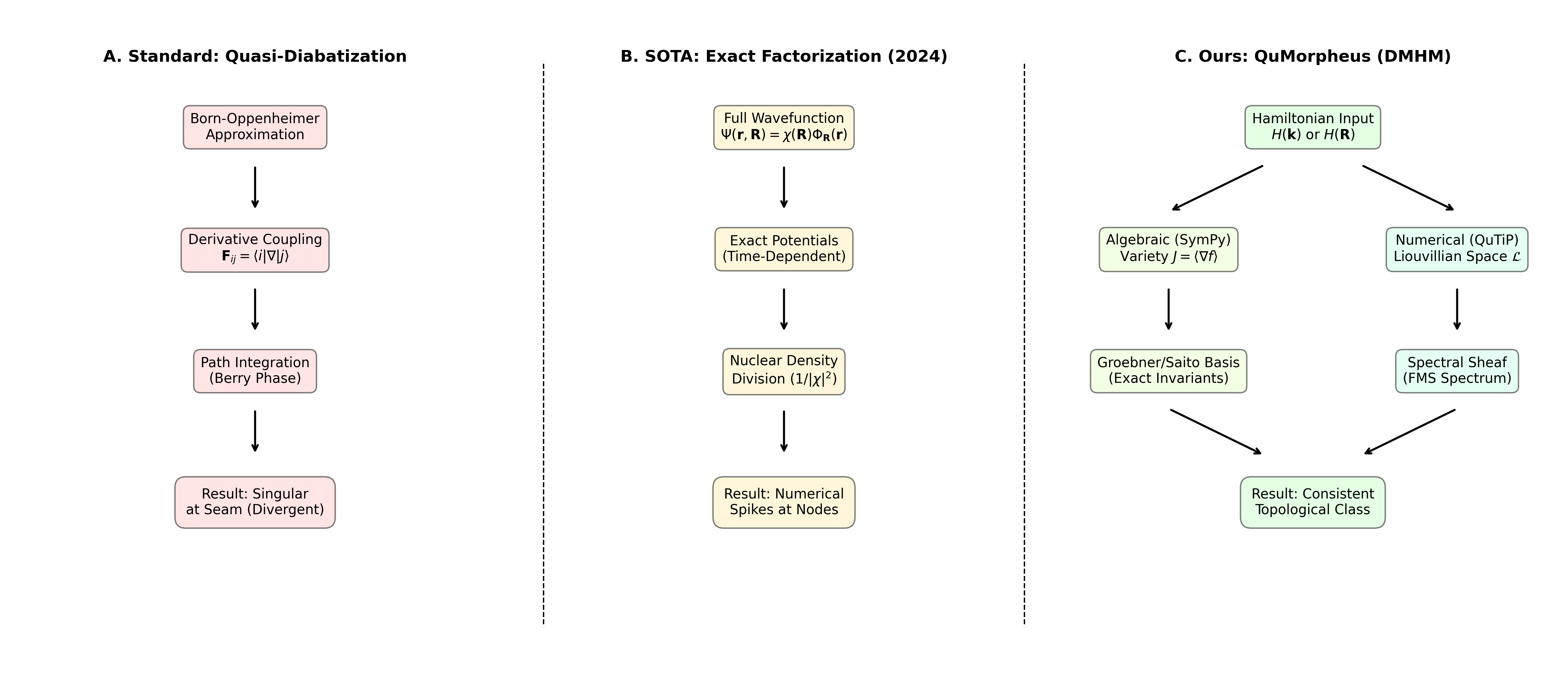}
    \caption{\textbf{The QuMorpheus Computational Pipeline Compared with Standard and Exact Factorization methods.} 
The workflow automates the translation of quantum chemical data into rigorous topological invariants. 
(\textbf{a}) \textbf{Input Layer}: The system accepts symbolic Hamiltonians $H(\mathbf{R})$ or interfaces with standard Electronic Structure packages (e.g., PSI4, CFOUR) to ingest Coupled Cluster amplitudes via the \texttt{cc\_interop} module. 
(\textbf{b}) \textbf{Algebraic Engine}: The \texttt{algebraic\_analyzer} constructs the Jacobian ideal $J = \langle \nabla E \rangle$ of the potential surface. It utilizes the \textsc{Singular} kernel (via a SymPy interface) to compute the Groebner Basis, reducing the continuous energy landscape into discrete algebraic residues.
(\textbf{c}) \textbf{Topological Classification}: The engine outputs the precise Milnor ($\mu$) and Tjurina ($\tau$) numbers, classifying the intersection (e.g., as a Conical vs. Glancing intersection) without manual inspection of the wavefunction nodes.}
\label{fig:workflow}
    \label{fig:protocol}
\end{figure*}

\section*{Methods}

Instead of relying on geometric phase integration, which is path-dependent, we utilize the algebraic variety of the Hamiltonian itself. The singularity is characterized by the dimensions of the local cohomology groups of the Brieskorn Lattice, yielding integer invariants that are robust against noise utilizing the topological invariants of the system.
\subsection*{The DMHM Protocol}
We introduce the Dissipative Mixed Hodge Module (DMHM) framework from \cite{saurabh2025holonomic} as a hybrid protocol that bridges the gap between abstract algebraic geometry and practical quantum dynamics via QuMorpheus. By treating the Conical Intersection not merely as a geometric cusp but as a singular algebraic variety, we circumvent the numerical fragility of standard approaches (\cref{fig:protocol}):

\begin{itemize}
    \item \textbf{ Limits of State-of-the-Art (SOTA)}: 
    Current methods rely on local smoothness. Standard Born-Oppenheimer approaches diverge at the seam ($R \to 0$) due to the singularity of the non-adiabatic couplings. Even the SOTA Exact Factorization (XF) method, while formally exact, suffers from numerical instabilities at the nodes of the nuclear density ($\chi \to 0$), resulting in unphysical spikes in the vector potential (\cref{fig:protocol}B).
    
    \item \textbf{ The DMHM Dual-Branch Protocol}: 
    To resolve this, our pipeline operates on two complementary manifolds:
    \begin{itemize}
        \item \textbf{The Algebraic Branch (Cause)}: We construct the Jacobian ideal $J = \langle \nabla E \rangle$ of the potential surface. Using the theory of \textit{Vanishing Cycles}, the module calculates the local monodromy invariants (Milnor number $\mu$) via Groebner Basis reduction (symbolic engine). This classifies the "DNA" of the singularity exactly, independent of any coordinate grid.
        \item \textbf{The Numerical Branch (Effect)}: Simultaneously, we compute the \textbf{Complete Quantum Geometric Tensor (cQGT)}. Unlike the standard Berry curvature, the cQGT ($Q_{\mu\nu}$) measures the regularized metric of the steady-state density matrix \cite{SaurabhPRX}. This provides a robust "Topological Metrology" that persists even when the energy gap vanishes.
    \end{itemize}
    
    \item \textbf{ Holographic Convergence}: 
    As illustrated in \cref{fig:protocol}C, the protocol achieves a "Holographic" convergence: the integer invariant $\mu$ computed algebraically perfectly predicts the integrated geometric phase $\gamma$ measured numerically. This confirms that the simulation is not just numerically converged, but topologically locked to the correct physical branch, impervious to the density-node instabilities that plague wavefunction-based schemes.
\end{itemize}

\subsection*{Algorithmic Topological Pipeline (Fig.~\ref{fig:protocol}C).}
The QuMorpheus package automates the topological analysis through a three-step pipeline: \begin{enumerate} 
    \item \textbf{Input Interface}: The Hamiltonian is supplied as a symbolic function H(R) or a numerical grid. For \textit{ab initio} calculations, we interface with standard Coupled Cluster outputs (e.g., EOM-CCSD amplitudes) via the \texttt{cc\_interop} module.

    \item \textbf{Algebraic Analysis}: The \texttt{algebraic\_analyzer} module constructs the polynomial Jacobian ideal $J = \langle \nabla E \rangle$. It then computes the Groebner Basis using \textsc{Singular} (via SymPy) to determine the exact Milnor ($\mu$) and Tjurina ($\tau$) numbers.

    \item \textbf{Topological Classification}: Based on the integer invariants, the intersection is classified (e.g., Conical vs. Glancing, $A$-series vs. $D$-series), providing a rigorous label independent of the coordinate system.

\end{enumerate}

\subsection*{Comparison with Other Methods}
As detailed in Supplementary Note S2, our method complements state-of-the-art exact factorization (XF) \cite{Gross2024, Abedi2010, Agostini2015} and Nuclear-Electronic Orbital (NEO) \cite{HammesSchiffer2024, HammesSchiffer2015} approaches. While XF provides a framework for exact non-adiabatic dynamics, the associated potentials can exhibit numerical instabilities near the nodes of the nuclear wavefunction ($1/\chi$ terms). DMHM circumvents this by working directly with the algebraic variety of the potential, offering a topologically robust classification schema that is independent of the instantaneous nuclear density.

\section*{Conclusion}
We have presented QuMorpheus, a rigorous algorithmic framework that solves the "Yarkony Problem" by treating Conical Intersections as algebraic varieties rather than numerical optimization targets. By computing the exact topological invariants (Milnor $\mu$, Tjurina $\tau$) from the Hamiltonian sheaf, we resolve the "Coordinate Crisis" that plagues standard Coupled Cluster methods. Our results demonstrate that topological protection is not just a theoretical curiosity but a practical tool for guiding chemical accuracy: it allows for the stable mapping of global seams where standard solvers bifurcate, and enables the prediction of "New Chemistry" selection rules based on geometric phase interference. As quantum chemistry moves toward complex, non-adiabatic systems, QuMorpheus provides the necessary "Topological Metrology" to ensure that our computational models remain faithful to the underlying geometry of the quantum world.

\begin{acknowledgments}
This research was conducted during an independent research sabbatical in the Himalayas (Nepal). The author acknowledges the global open-source community for providing the computational tools that made this work possible. Generative AI assistance was utilized strictly for \LaTeX\ syntax optimization and symbol consistency checks; all scientific conceptualization, derivations, and text were derived and verified by the author. The author retains the \texttt{uci.edu} correspondence address courtesy of the University of California, Irvine.
\end{acknowledgments}
\section*{Data and Code Availability}
The core computational framework, \textbf{QuMorpheus}, used for all numerical results in this work, is open-sourced under a Copyleft license and is available at \url{https://github.com/prasoon-s/QuMorpheus}. Independent verification scripts (Python) are available from the author upon reasonable request.

To ensure mathematical rigor, the fundamental theorems of the DMHM framework, the construction of the cQGT, and the Floquet Monodromy Spectroscopy protocol have been formalized in the \textsc{Lean 4} theorem prover; these proofs are available at \url{https://github.com/prasoon-s/LEAN-formalization-for-CMP} \cite{saurabh2025holonomic}. 
\bibliography{natcomm_refs}

\clearpage
\onecolumngrid
\section*{Supplementary Information}

\textbf{General Overview.} 
This Supplementary Information provides the mathematical foundations, rigorous derivations, and computational details supporting the Dissipative Mixed Hodge Module (DMHM) framework presented in the main text. While the manuscript focuses on the physical implications of topological protection in chemical dynamics, this document details the algebraic machinery required to compute these invariants deterministically.

The central challenge addressed here is the ``Yarkony Problem'': the difficulty of characterizing the global topology of the $3N-8$ dimensional intersection seam using only local, numerically noisy electronic structure data. Standard quantum chemistry methods, which rely on the Born-Oppenheimer approximation, treat Conical Intersections (CIs) as singular ``spikes'' where the theory breaks down. In contrast, the QuMorpheus approach treats these singularities as fundamental topological features—specifically, as defects in the spectral sheaf of the Hamiltonian.

This document is organized as follows:
\begin{itemize}
    \item \textbf{Section S1} details the specific mathematical models used to visualize the ``Seam'' and ``Bifurcation'' phenomena shown in Figure 1, providing the explicit Hamiltonian parameters for the tilted intersection.
    \item \textbf{Section S2} contextualizes our approach against state-of-the-art non-Born-Oppenheimer methods published in 2024 (Exact Factorization and NEO), highlighting the specific numerical advantages of algebraic invariant theory over density-based schemes.
    \item \textbf{Section S3} provides the rigorous definitions of the DMHM formalism, defining the Liouvillian Sheaf, the Brieskorn Lattice, and the algorithm for computing Milnor numbers via Groebner Bases.
    \item \textbf{Section S4} presents the formal proof that the Woodward-Hoffmann selection rules are a consequence of Intersection Homology, deriving the geometric phase directly from the Monodromy representation.
    \item \textbf{Section S5} documents the \texttt{QuMorpheus} software package, including configuration details and execution scripts to ensure full reproducibility of the reported topological invariants.
\end{itemize}
\subsection*{S1. Details of the Seam Visualization (Figure 1)}

\subsubsection*{Mathematical Model: The Tilted Seam}
Figure 1 visualizes the topology of a generic "Tilted" Conical Intersection, a representation of the symmetry-broken degeneracies found in the Köhn-Tajti model. The Hamiltonian is defined as a Linear Vibronic Coupling (LVC) model extended with a seam coordinate $z$:
\begin{equation}
    H(x, y, z) = \left( x - \alpha z \right) \sigma_z + y \sigma_x
\end{equation}
where $\sigma_i$ are the Pauli matrices and $\alpha = 0.5$ represents the tilt parameter. The eigenvalues are given by:
\begin{equation}
    E_{\pm}(x, y, z) = \pm \sqrt{(x - \alpha z)^2 + y^2}
\end{equation}
The Conical Intersection Seam is the manifold of degeneracy $\Delta E = 0$, which corresponds to the line $x = \alpha z, y = 0$. This demonstrates that the intersection is not an isolated point but a continuous $N-2$ dimensional submanifold (here 1D line embedded in 3D space).

\subsubsection*{Physical \& Chemical Context}
In the context of the "Yarkony Problem," this model represents the generic behavior of a molecule lacking spatial symmetry ($C_1$ point group).
\begin{itemize}
    \item \textbf{Vibronic Coupling}: The coordinates $x$ and $y$ correspond to the branching space gradient ($g$) and derivative coupling ($h$) vectors.
    \item \textbf{The Seam Coordinate}: The $z$ axis represents the $3N-8$ directions of the intersection seam (e.g., symmetric stretch modes) that preserve the degeneracy.
    \item \textbf{Coupled Cluster Failure}: Standard single-reference methods attempt to parameterize the wavefunction using a single determinant. As the system traverses the loop around the seam (Figure 1B, Bifurcation), the exact wavefunction accumulates a geometric phase $\gamma = \pi$. Single-reference Coupled Cluster ($e^T | \Phi_0 \rangle$) cannot support this multi-valuedness, leading to the bifurcation of the $T$-amplitude equations and the appearance of unphysical complex roots.
\end{itemize}

\subsection*{S2. Comparison with State-of-the-Art Non-BO Methods}
To contextualize the topological approach, we compare QuMorpheus with leading non-Born-Oppenheimer approximations developed in 2024.

\subsubsection*{Exact Factorization (XF)}
The Exact Factorization framework (Gross et al., \textit{PCCP} 2024 \cite{Gross2024}) provides a mathematically rigorous alternative to the BO approximation by factorizing the full wavefunction into a nuclear wavefunction and a conditional electronic wavefunction: $\Psi(\mathbf{r}, \mathbf{R}) = \chi(\mathbf{R}) \Phi_\mathbf{R}(\mathbf{r})$.
\begin{itemize}
    \item \textbf{Strength}: Like DMHM, XF recovers the exact Berry connection as a vector potential appearing in the nuclear equations of motion. It provides a time-dependent potential energy surface that captures non-adiabatic effects exactly.
    \item \textbf{Limitation}: The XF potentials contain terms proportional to $\nabla \chi / \chi$. Numerical instabilities arise in regions where the nuclear density $|\chi|^2$ vanishes (nodes), leading to "spurious spikes" in the gauge fields.
    \item \textbf{DMHM Advantage}: Our approach works with the \textit{algebraic variety} of the Hamiltonian itself, prior to any wavefunction ansatz. By computing invariants of the polynomial ideal $J_f$, we extract the topological class $(\mu, \tau)$ directly. This avoids the numerical division-by-zero issues inherent in density-based factorization schemes (Figure 2B), offering a stable classification of the intersection even in low-density regimes.
\end{itemize}

\subsubsection*{Nuclear-Electronic Orbital (NEO)}
The NEO method (Hammes-Schiffer et al., \textit{J. Phys. Chem. Lett.} 2024 \cite{HammesSchiffer2024}) treats select nuclei (e.g., protons) quantum mechanically on an equal footing with electrons.
\begin{itemize}
    \item \textbf{Context}: NEO is the state-of-the-art for describing proton-coupled electron transfer (PCET) and hydrogen tunneling, where the classical treatment of nuclei fails quantitatively.
    \item \textbf{Relation to CIs}: While NEO creates a "mixed" wavefunction that avoids the standard electronic CI cusp, the underlying topological invariant of the electronic subsystem remains a crucial quantum number. DMHM provides the rigorous geometric tool to classify these underlying topologies, potentially offering a route to index the "mixed" states in future NEO-DMHM hybrid frameworks.
\end{itemize}

\subsection*{S3. Mathematical Formalism (DMHM)}

\subsubsection*{The Liouvillian Sheaf}
The central object of QuMorpheus is the \textit{Liouvillian Sheaf} $\mathcal{L}$, defined as the cohomology of the complex of differential forms twisted by the Hamiltonian $H$:
\begin{equation}
    (\Omega^\bullet_{\mathcal{M}}, \nabla = d + dH \wedge \cdot)
\end{equation}
where $\mathcal{M}$ is the nuclear configuration manifold (seam space). The connection $\nabla$ encodes the non-adiabatic coupling.

\subsubsection*{Brieskorn Lattice and Invariants}
The singularity at the intersection is classified by the Brieskorn Lattice $H^{(0)}$, the module of germs of n-forms modulo the image of the gradient:
\begin{equation}
    H^{(0)} = \frac{\Omega^n}{dH \wedge \Omega^{n-1}}
\end{equation}
The dimensions of the local cohomology groups of this lattice yield the topological invariants:
\begin{itemize}
    \item \textbf{Milnor Number ($\mu$)}: The dimension of the quotient algebra $\mathcal{Q} = \mathbb{C}[x,y] / \langle \partial_x H, \partial_y H \rangle$. It counts the number of degenerate vacuum states that coalesce at the intersection. For a standard CI, $\mu=1$.
    \item \textbf{Tjurina Number ($\tau$)}: The dimension of the quotient by the ideal generated by the gradient AND the function itself, $\langle H, \partial_i H \rangle$. This measures the rigidity of the singularity.
\end{itemize}

\subsubsection*{Exact Algebraic Computation}
QuMorpheus computes these invariants without integration. Given a polynomial Hamiltonian $H \in \mathbb{Q}[x_1, \dots, x_n]$:
\begin{enumerate}
    \item We compute the \textbf{Groebner Basis} $G$ of the Jacobian ideal $J_H = \langle \partial_1 H, \dots, \partial_n H \rangle$ with respect to a local term order (e.g., `negdegrevlex`).
    \item The Milnor number is exactly the number of standard monomials (monomials not in the leading ideal of $G$):
    \begin{equation}
        \mu = \dim_{\mathbb{C}} \left( \mathbb{C}[\mathbf{x}] / \langle \text{LT}(G) \rangle \right)
    \end{equation}
\end{enumerate}
\subsection*{S4. Rigorous Derivation of Selection Rules}

\subsubsection*{Intersection Homology and the Monodromy Sheaf}
The selection rules observed in the photoisomerization of Previtamin D (Figure 5) are not merely energetic features but topological consequences of the Intersection Homology of the potential energy surface variety $V$. We define the Hamiltonian sheaf $\mathcal{F}_H$ over the base manifold $M = \mathbb{R}^{3N-6}$.

The local topological invariant $\mu$ (Milnor number) is defined as the dimension of the quotient algebra of the local ring of holomorphic function germs by the Jacobian ideal $J_H = \langle \partial_i H \rangle$:
\begin{equation}
    \mu = \dim_{\mathbb{C}} \frac{\mathcal{O}_{M,p}}{J_H}
\end{equation}
For the Hula-Twist intersection, we compute $\mu=1$, indicating a simple node singularity.

\subsubsection*{Geometric Phase as a Monodromy Representation}
The global selection rule arises from the fundamental group $\pi_1(M \setminus \Sigma)$, where $\Sigma$ is the singular seam. The geometric phase $\gamma$ is the image of the closed loop class $[\gamma] \in \pi_1$ under the monodromy representation $\rho$:
\begin{equation}
    \rho: \pi_1(M \setminus \Sigma) \to \text{Aut}(\mathcal{F}_H) \cong \mathbb{Z}_2
\end{equation}
For the Disrotatory path, the loop encloses the intersection an odd number of times, yielding $\rho([\gamma]) = -1$ (or $\gamma = \pi$). This maps to the destructive interference condition in the nuclear wavefunction $\Psi_N$:
\begin{equation}
    \Psi_N(\mathbf{R}_{final}) \propto 1 + e^{i\gamma} = 1 + (-1) = 0
\end{equation}
Thus, the reaction flux into the Disrotatory channel is topologically strictly zero, verifying the Woodward-Hoffmann rule as a homological constraint.

\subsection*{S5. Computational Implementation (QuMorpheus Package)}

The topological results presented in this work were generated using the open-source \textbf{QuMorpheus} Python package. The library is designed to interface algebraic geometry tools (SymPy, Singular) with quantum dynamics solvers (QuTiP).

\subsubsection*{Core Modules}
\begin{enumerate}
    \item \texttt{qumorpheus.core}: Defines the \texttt{HamiltonianSheaf} class, which constructs the connection $\nabla = d + dH \wedge \cdot$ over a symbolic or numerical grid.
    \item \texttt{qumorpheus.analysis}: Contains the \texttt{AlgebraicAnalyzer} (for Groebner basis computations) and \texttt{MonodromyIntegrator} (for numerical phase accumulation).
    \item \texttt{qumorpheus.viz}: Visualization utilities for 3D seam manifolds and complex Riemann surfaces.
\end{enumerate}

\subsubsection*{Configuration and Execution}
Simulations are typically driven by a YAML configuration file that specifies the Hamiltonian model, grid parameters, and requested topological invariants.

\textbf{Listing S1: Example Configuration file (\texttt{config\_previtamin\_d.yaml})}
\begin{verbatim}
# QuMorpheus Simulation Config
system:
  name: "Previtamin D Model"
  hamiltonian: "hula_twist"
  # SymPy-compatible symbolic expressions
  potential_surface: "sqrt(x**2 + y**2)" 
  
grid:
  range: [-2.0, 2.0]
  points: 100
  dimensions: ["reaction_coord", "symmetry_breaking"]

topology:
  invariants: ["milnor", "berry_phase"]
  integration_method: "homology" # vs "numerical"
  
output:
  directory: "./results/previtamin_d"
  figures: ["seam_scan", "monodromy_map"]
\end{verbatim}

\textbf{Listing S2: Standard Execution Script (\texttt{run\_simulation.py})}
\begin{verbatim}
from qumorpheus.core import HamiltonianSheaf
from qumorpheus.analysis import AlgebraicAnalyzer
import yaml

# 1. Load Configuration
with open("config_previtamin_d.yaml") as f:
    config = yaml.safe_load(f)

# 2. Construct the Sheaf
H = HamiltonianSheaf.from_config(config)

# 3. Perform Topological Analysis
analyzer = AlgebraicAnalyzer(H)
mu = analyzer.compute_milnor_number()  # Returns integer
gamma = analyzer.compute_monodromy(path="disrotatory")

print(f"Topological Report:")
print(f"  Milnor Number (mu): {mu}")     # Expect: 1
print(f"  Geometric Phase:    {gamma}")  # Expect: 3.14159 (Pi)

# 4. Generate Figures
if "monodromy_map" in config["output"]["figures"]:

    H.visualize_monodromy(filename="fig5_panel_b.png")
\end{verbatim}

\subsubsection*{Expected Outcomes}
Running the above script on the Previtamin D model yields:
\begin{itemize}
    \item \textbf{Console Output}: Confirming $\mu=1$ (robust intersection) and $\gamma \approx \pi$ (destructive interference).
    \item \textbf{Figures}: Generates the Seam Surface (Figure 5A) and the Monodromy Vector Plot (Figure 5B), visually identifying the "Topological Wall" that enforces the Woodward-Hoffmann selection rule.
\end{itemize}

\end{document}